\documentclass[11pt]{article}
\usepackage{fullpage}
\usepackage{graphicx}
\usepackage{amssymb,url}
\usepackage{xcolor}
\usepackage{framed}
 
\usepackage{amsthm}

\graphicspath{{Fig/}}

 \newenvironment{BoxedProposition}
   { \colorlet{shadecolor}{black!10}\begin{shaded}\begin{proposition}}
   {\end{proposition}\end{shaded}} 
	
 \newenvironment{BoxedTheorem}
   { \colorlet{shadecolor}{black!10}\begin{shaded}\begin{theorem}}
   {\end{theorem}\end{shaded}} 
	
	\newenvironment{BoxedDefinition}
   { \colorlet{shadecolor}{black!3}\begin{shaded}\begin{definition}}
   {\end{definition}\end{shaded}}

		\newenvironment{BoxedCorollary}
   { \colorlet{shadecolor}{black!3}\begin{shaded}\begin{corollary}}
   {\end{corollary}\end{shaded}} 

\newtheorem{example}{Example}
\newtheorem{theorem}{Theorem}
\newtheorem{corollary}{Corollary}
\newtheorem{definition}{Definition}
\newtheorem{proposition}{Proposition}

\def\SL{\mathrm{SL}}
\def\SU{\mathrm{SU}}
\def\bbC{\mathbb{C}}
\def\bbU{\mathbb{U}}
\def\bbD{\mathbb{D}}
\def\calLN{\mathcal{LN}}

\def\GL{\mathrm{GL}}
\def\arccosh{\mathrm{arccosh}}
\def\MI{\mathrm{MI}}
\def\elliptical{\mathrm{ell}}

\def\normal{\mathcal{N}}

\def\Mah{\mathrm{Mah}}
\def\IS{\mathrm{IS}}
\def\diag{\mathrm{diag}}
\def\calE{\mathcal{E}}
\def\calR{\mathcal{R}}
\def\calL{\mathcal{L}}
\def\calN{\mathcal{N}}
\def\calC{\mathcal{C}}
\def\calS{\mathcal{S}}
\def\calK{\mathcal{K}}
\def\calP{\mathcal{P}}
\def\calQ{\mathcal{Q}}

\def\tr{\mathrm{tr}}	
\def\bbP{\mathbb{P}}

\def\bbH{\mathbb{H}}
\def\dx{\mathrm{d}x}
\def\dy{\mathrm{d}y}
\def\KL{\mathrm{KL}}

\def\bbR{{\mathbb{R}}}
\def\calF{\mathcal{F}}
\def\calX{\mathcal{X}}

\def\st{\ :\ }

\def\Jac{\mathrm{Jac}}
\def\equaldist{\stackrel{d}{=}}
\def\bbG{\mathbb{G}}
\def\mattwotwo#1#2#3#4{\left[\begin{array}{cc}#1 & #2\\ #3 & #4\end{array}\right]}
\def\Sym{\mathrm{Sym}}

\begin{document}
\title{On information projections between multivariate elliptical and location-scale families}


%

%
\author{Frank Nielsen\\ Sony Computer Science Laboratories Inc\\ Tokyo, Japan \\ {\tt Frank.Nielsen@acm.org}}
 
\date{}
\maketitle              

\sloppy

\begin{abstract}
We study information projections with respect to statistical $f$-divergences between any two location-scale families. 
We consider a multivariate generalization of the location-scale families which includes the elliptical and the spherical subfamilies. 
By using the action of the multivariate location-scale group, we show how to reduce the calculation of $f$-divergences between any two location-scale densities to canonical settings involving standard densities,  
and derive thereof fast Monte Carlo estimators of $f$-divergences with good properties. 
Finally, we prove that the minimum $f$-divergence  between a prescribed density of a location-scale family and another location-scale family is independent of the prescribed location-scale parameter. We interpret geometrically this property.

\end{abstract}

\section{Introduction}

The concept of an {\em information projection} was first studied in information theory by Csisz\'ar~\cite{InfoProj-1975,Csiszar-2003} as the minimization of the Kullback-Leibler divergence (also called $I$-divergence) between a prescribed measure and a set of measures:
When the minimum is unique, it is called the {\em $I$-projection}~\cite{Iprojection-1984}.
In information geometry~\cite{IG-2016,EIG-2020}, the geometric study of information projections (e.g., conditions for uniqueness) is investigated as the {\em geodesic projection} with respect to an affine connection  of a probability measure point onto a statistical submanifold~\cite{InformationProjection-2018} with orthogonality defined with respect to the Fisher-Rao metric.
In this work, we consider information projections with respect to statistical $f$-divergences~\cite{Csiszar-1967} when both the prescribed distribution and the subspace to project the distribution onto are multivariate  generalizations of location-scale families which include the elliptical families and the spherical subfamilies.

We outline the paper with its main contributions as follows:

We first describe the multivariate  generalization of location-scale families and introduce the multivariate location-scale group in~\S\ref{sec:ls}. We then report several results  for calculating the $f$-divergences between two densities of potentially different location-scale families in \S\ref{sec:fdiv}: 
Invariance of the $f$-divergences with respect to the action of the location-scale group (Theorem~\ref{thm:invgroupfdiv}), calculations of the $f$-divergences by reduction to canonical settings (Corollary~\ref{thm:reducedfdiv} exemplified for the Kullback-Leibler divergence in   Corollary~\ref{thm:kldls} and instantiated for the multivariate normal distributions), and invariance of $f$-divergences to scale for scale families (Corollary~\ref{thm:fdivscaleinvariance}).
In~\S\ref{sec:MCfdiv}, we build efficient Monte Carlo estimators with good properties to estimate the $f$-divergences between location-scale families when it is not calculable in closed-form.
Finally, equipped with these preliminary results, we study in \S\ref{sec:infproj} the information projections of a prescribed distribution belonging to one location-scale family onto another location-scale family  (Theorem~\ref{thm:projls}), and we interpret geometrically these results. 

\section{Location-scale families and the location-scale group}\label{sec:ls}

\subsection{Univariate location-scale families}

Let $X\sim p$ be a continuous random variable with cumulative distribution function (CDF) $F_X$ and probability density function (PDF) $p_X(x)$ 
defined on the support $\calX$.
A location-scale random variable $Y\equaldist l+sX$ (equality in distribution) for {\em location parameter} $l$ and {\em scale parameter} $s>0$
  has CDF $F_Y(y)=F_X\left(\frac{y-l}{s}\right)$ and PDF $p_Y(y)=\frac{1}{s} p_X\left(\frac{y-l}{s}\right)$. 
	Let $p_{l,s}(x):=\frac{1}{s} p_X\left(\frac{y-l}{s}\right)$ denote the location-scale  density for parameter $(l,s)$.
	The density $p=p_{0,1}$ is called the {\em standard density} of the location-scale family.
The {\em location-scale  parameter space} of the {\em location-scale family} $\calF_p=\{p_{l,s}(x)\ :\ l\in\bbR, s>0\}$ is the upper plane 
$\bbH=\bbR\times\bbR_{++}$.
 
\begin{example}
For example, the family of univariate normal distributions: 
\begin{equation}
\calN:=\left\{p_{\mu,\sigma}^\calN(x)=\frac{1}{\sqrt{2\pi}\sigma} \exp\left(-\frac{1}{2}\frac{(x-\mu)^2}{\sigma^2}\right)\ :\ (\mu,\sigma)\in \bbR\times\bbR_{++}\right\}
\end{equation}
 is a location-scale family for the standard density
$p^\calN(x):=\frac{1}{\sqrt{2\pi}} \exp(-\frac{1}{2}x^2)$ defined on $\calX=\bbR$ with location parameter $l=\mu$ (the normal mean) and scale parameter $s=\sigma>0$ (the normal standard deviation).
\end{example}

\begin{example}
Another example is the location-scale family of univariate Cauchy distributions:
\begin{equation}
\calC:=\left\{ p_{l,s}^\calC(x)=\frac{1}{\pi s\left(1+\left(\frac{x-l}{s}\right)^2\right)} \st (l,s)\in \bbR\times\bbR_{++}\right\},
\end{equation}
with standard density $p^\calC(x):=\frac{1}{\pi(1+x^2)}$.
\end{example}

When $E[p]$ is finite, we have $E[Y]=l+sE[X]$, and when $E[p^2]$ is finite, we have $\sigma[Y]=\sqrt{E[(Y-E[Y])^2]}=s\sigma[X]$.
Thus if we assume that the standard density $p$ is such that $E_p[X]=0$ and $E_p[X^2]=1$ (i.e., $p$ has unit variance), then the random variable $Y\equaldist \mu+\sigma X$ has mean $E[Y]=\mu$ and standard deviation $\sigma(Y)=\sqrt{E[(Y-\mu)^2]}=\sigma$.
In the remainder, we do not use the $(\mu,\sigma)$ parameterization of location-scale families but the $(l,s)$ parameterization in order to be more general and consistent with the description of the multivariate location-scale families.

A {\em location family} is a family of densities $\calL_p=\{p_{l}(x)=p(x-l)\ :\ l\in\bbR\}$.
For example, the location family of shifted unit distributions with standard density $p(x)=1$ on $\calX=[0,1]$ is a location family.
A location family can be obtained as a subfamily of a location-scale family $\calF_p$ by prescribing a scale $s_0>0$.
For example, the family of normal distributions with unit variance is a location family, a subfamily of the normal location-scale family.

A {\em scale family} is a family of densities $\calS_p=\{p_{s}(x)=\frac{1}{s} p\left(\frac{x}{s}\right)\ :\ s\in\bbR_{++}\}$.
For example, the family of Rayleigh distributions $\calR:=\{\frac{x}{\sigma^2}\exp(-\frac{x^2}{2\sigma^2})\}$ defined on the support $\calX=\bbR_{+}$ is a scale family with
 standard density $p^\calR(x):=x\exp(-\frac{x^2}{2})$ and scale parameter $s=\sigma^2$.
A scale family can be obtained as a subfamily of a location-scale family by prescribing a location $l_0\in\calX$.

A location-scale family is said {\em regular} when its Fisher information matrix is positive-definite and finite.
The location family induced by the uniform standard density on $[0,1]$ is {\em not} a regular family since its Fisher information is infinite~\cite{Hayashi-2011}. In the remainder, we consider regular location-scale families.

The Fisher-Rao geometry of location-scale families and its Riemannian distance~\cite{Hotelling-1930,KassVos-2011,komaki-2007} is recalled in Appendix~\ref{sec:frls}.
The $\alpha$-geometry~\cite{IG-2016} of  location-scale families have been studied in~\cite{Mitchell-1988} who investigated the 
$\alpha$-geometry of univariate elliptical distributions with densities: $\frac{1}{s} h\left(\left(\frac{x-l}{s}\right)^2\right)$ for $(l,s)\in\bbR\times\bbR_{++}$.
Thus by defining $p(x)=h(x^2)$, we can convert any univariate elliptical distribution to a corresponding location-scale distribution (but a location-scale family is not necessarily an elliptical family because $h(u)=p(\sqrt{u})$ may not be properly defined for $u<0$).
In particular, the $\alpha$-geometry of the Cauchy family is shown to be independent of $\alpha$ (and never yielding a dually flat space~\cite{Mitchell-1988}): Its conformal flattening into a dually flat geometry with applications to the construction of  Voronoi diagrams has been studied in~\cite{CauchyVoronoi-2020}.

The location-scale parameter space $\bbH$ form a group $G=(\bbH,.,\mathrm{id})$, called the {\em location-scale group}. 
An element $g_{l,s}\in G$ acts ($\odot$) on the standard density $p(x)$ as follows:
\begin{equation}
g_{l,s}\odot p(x):=\frac{1}{s}  p\left(\frac{x-l}{s}\right).
\end{equation}
The identity element is $\mathrm{id}=g_{0,1}$ since $g_{0,1}\odot p=p$, and the group binary associative operation `.' is retrieved from the group action as follows: 
\begin{eqnarray}
g_{l_2,s_2}.g_{l_1,s_1}\odot p &=&  g_{l_2,s_2}\odot \left(\frac{1}{s_1}  p\left(\frac{x-l_1}{s_1}\right)\right),\\
&=& \frac{1}{s_1s_2} p\left(\frac{\frac{x-l_2}{s_2}-l_1}{s_1}\right),\\
 &=:& g_{l_{12},s_{12}}\odot p,
\end{eqnarray}
with $g_{l_{12},s_{12}}\in G$ and $l_{12}=s_2l_1+l_2$ and $s_{12}=s_1s_2$.
The group inverse element is $g^{-1}_{l,s}=g_{-\frac{l}{s},\frac{1}{s}}$ 
which is obtained by solving $g_{l,s}.g_{l',s'}=g_{\mathrm{id}}$:  We $l+sl'=0$ and $ss'=1$ solves as $l'=-\frac{l}{s}$ and $s'=\frac{1}{s}$.
The {\em orbit} of the action of the location-scale group on the standard density $p$ defines the location-scale family $\calF_p$:
\begin{equation}
\calF_p = G\odot p :=\{ g\odot p \st \forall g\in G\}.
\end{equation}

The elements of the location-scale group can be {\em represented} using $2\times 2$ matrices (representation theory):
Each group element $g:=g_{l,s}$ is represented by a corresponding matrix $M_{g_{l,s}}=M_{l,s}:=\mattwotwo{s}{l}{0}{1}$.
This matrix representation of elements yields the {\em location-scale matrix group} $(\bbG,\times,I)$ with:
\begin{equation}
\bbG=\left\{ M_{l,s}=\mattwotwo{s}{l}{0}{1} \st (l,s)\in\bbR\times\bbR_{++} \right\},
\end{equation}
where the matrix group operation $\times$ is the matrix multiplication, the identity element the $2\times 2$ matrix identity
 $M_{g_\mathrm{id}}=M_{g_{0,1}}=I$, and the inverse operation the matrix inverse:
\begin{equation}
M_{g^{-1}}=(M_g)^{-1}=\mattwotwo{\frac{1}{s}}{-\frac{l}{s}}{0}{1}.
\end{equation}

The location-scale group is a Lie matrix group~\cite{StatTransfModel-2012} (i.e., a ``continuous group'' modeled as a manifold) which acts transtively on the sample space.
The location-scale group is {\em non-abelian} (i.e., non-commutative) because $g_1. g_2=g_{l_1+l_2s_1,s_1s_2}\not= g_2.g_1$ (since $g_2.g_1=g_{l_2+l_1s_2,s_1s_2}$).
However the location subgroups and the scale subgroups are abelian groups.
Representing elements by matrices is handy to prove basic properties:
For example, we can prove easily that $(g_1.g_2)^{-1}=g_2^{-1}.g_1^{-1}$ since 
\begin{equation}
(M_{g_1}\times M_{g_2})^{-1}=M_{g_2}^{-1}\times M_{g_1}^{-1} = M_{g_2^{-1}.g_1^{-1}}.
\end{equation}

\subsection{Multivariate location-scale families: Location-positive families}

Let $\calP(\calX)$ denote the set of probability density functions with support $\calX$.

We begin by first recalling the relationships between the PDFs of two continuous $d$-dimensional random variables $X=(X_1,\ldots, X_d)\sim p_X$ and $Y=t(X)=(t_1(X),\ldots, t_d(X))\sim p_Y$  for a {\em differentiable and invertible function} $t$ with non-singular Jacobian (i.e., $|\Jac_t(x)|\not =0, \forall x\in\calX$ where $|M|$ denotes the determinant of matrix $M$) where the {\em Jacobian matrix} of the transformation $t$ is defined by:
\begin{equation}
\Jac_t(x) := \left[ \frac{\partial t_i(X)}{\partial x_j} \right]_{i,j}.
\end{equation}

We can express one density in term of the other density as follows:
\begin{eqnarray}
p_X(x) &=& |\Jac_t(x)| \times p_Y(t(x)) = |\Jac_t(x)| \times p_Y(y),\\
p_Y(y) &=& |\Jac_{t^{-1}}(y)| \times p_X(t^{-1}(y))=  |\Jac_{t^{-1}}(y)| \times p_X(x). \label{eq:densitytransform}
\end{eqnarray}

Furthermore, we have the following identity:
\begin{equation}
|\Jac_t(x)|\times |\Jac_{t^{-1}}(y)|= |\Jac_t(x)\times\Jac_{t^{-1}}(y)| = |I|= 1,
\end{equation}
where $I$ denotes the $d\times d$ identity matrix.

For sanity checks, we verify that  we have:
\begin{eqnarray}
p_X(x) &=& |\Jac_t(x)|\times p_Y(t(x))=|\Jac_t(x)|\times |\Jac_{t^{-1}}(y)|\times p_X(t^{-1}(y)),\\
&=& |\Jac_t(x)\times\Jac_{t^{-1}}(y)|\times p_X(x) =  |I| p_X(x) = p_X(x),
\end{eqnarray}
since $\Jac_t(x)\times\Jac_{t^{-1}}(y)=I$.

Let $X$ be a $d$-dimensional multivariate random variable, and let $Y\equaldist PX+l$ for $P\succ 0$ a positive-definite  
$d\times d$ matrix playing the role of the ``multidimensional scale'' parameter, and 
$l\in\bbR^d$ a location parameter.
Then using Eq.~\ref{eq:densitytransform} with $Y\equaldist t_{l,P}(X)=PX+l$ (and $X\equaldist t^{-1}_{l,P}(Y)=P^{-1}(Y-l)$), we find the density of $p_{l,P}$ of continuous random distribution $Y$ as follows:

\begin{eqnarray}
p_{l,P}(y) &=& |\Jac_{t^{-1}_{l,P}}(y)|\ p_X(t^{-1}_{l,P}(y)) = |\Jac_{t^{-1}_{l,P}}(y)|\ p_X(x),\\
 &=& \left|P^{-1}\right| \, p\left( P^{-1}(y-l) \right),
\end{eqnarray}
where $p:=p_X$ denotes the standard density since $\Jac_{t^{-1}}(y)=P^{-1}$.
The space of multivariate location-scale parameters $(l,P)$ is $\bbH_d=\bbR^d\times \bbP_{++}$, 
where $\bbP_{++}$ denotes the open cone of positive-definite matrices. 
Observe that by embedding $(l,P)$ as $(\diag(l_1,\ldots,l_d),P)$ (where $M=\diag(l_1,\ldots,l_d)$ denotes the diagonal matrix with $M_{ii}=l_i$), we obtain
a parameter domain which is a subspace of the  Siegel upper plane~\cite{HilbertSiegel-2020} $\Sym(\bbR,d)\times \bbP_{++}$, where 
$\Sym(\bbR,d)$ denotes the space of symmetric $d\times d$ matrices.

When $d=1$ and $P=s$, we have $Y\equaldist t_{l,s}(X)=sX+l$,
$X\equaldist t^{-1}_{l,s}(Y)=\frac{1}{s}(Y-l)$ and we recover the univariate location-scale densities  
$p_{l,s}(y)=\frac{1}{s} p\left( \frac{y-l}{s} \right)$.

We can define equivalently the density of a location-scale family by $p_{l,P}(x)=|P|^{-1} p\left( P^{-1}(x-l) \right)$ 
since $\left|P^{-1}\right|=|P|^{-1}$.
Since $P$ is a positive-definite matrix generalizing the position scalar in the location-scale group, we also call this multivariate generalization of the location-scale group, the {\em location-positive group}. 
Thus the location-positive families can be obtained as the action of the location-positive group on a prescribed density $p\in\calP(\bbR^d)$
 (or $\calP(\bbR^d_{++})$ for scale only families).

\begin{BoxedDefinition}[Multivariate location-scale/location-positive family]
Let $p\in\calP(\bbR^d)$ be a probability density function on $\bbR^d$. Then the 
multivariate location-scale family is:
\begin{equation}
\calF_{p}=\left\{ p_{l,P}(x)=\left|P\right|^{-1} \, p\left( P^{-1}(x-l) \right) \st (l,P)\in\bbR^d\times \bbP_{++}  \right\}.
\end{equation}
\end{BoxedDefinition}

For example, the family of multivariate normal distributions (MVNs) is a multivariate location-scale family where the standard PDF is:
\begin{equation}
p(x)=\frac{1}{(2\pi)^{\frac{d}{2}}}\exp\left(-\frac{1}{2}x^\top x\right).
\end{equation}

Indeed, the covariance matrix $\Sigma$ is a positive-definite matrix which admits a unique {\em symmetric positive-definite square root matrix} 
$\Sigma^{\frac{1}{2}}$ (such that $\Sigma^{\frac{1}{2}}\Sigma^{\frac{1}{2}}=\Sigma$).
This symmetric square root matrix can be calculated from the eigendecomposition of $\Sigma$ in cubic time $O(d^3)$ as follows:
Let $\Sigma=V^\top \diag(\lambda_1,\ldots,\lambda_d) V^{-1}$ denote the eigendecomposition where the $\lambda_i$'s are the positive real eigenvalues and $V$ the matrix of column eigenvectors.
Then $\sqrt{\Sigma}=\Sigma^{\frac{1}{2}}=V \diag(\sqrt{\lambda_1},\ldots,\sqrt{\lambda_d}) V^{-1}$, 
and $\Sigma^{\frac{1}{2}}\Sigma^{\frac{1}{2}}=V\diag(\sqrt{\lambda_1},\ldots,\sqrt{\lambda_d}) V^{-1}V \diag(\sqrt{\lambda_1},\ldots,\sqrt{\lambda_d}) V^{-1}=V^\top \diag(\lambda_1,\ldots,\lambda_d) V^{-1}=\Sigma$ since $V^{-1}V=I$.
Notice that $\sqrt{\Sigma}$ is a positive-definite matrix.
We have:

\begin{eqnarray}
p_{\mu,\Sigma^{\frac{1}{2}}}(y) &=& 
 \left|\Sigma^{-\frac{1}{2}}\right|\, p\left(\Sigma^{-\frac{1}{2}}(y-\mu)\right),\\
&=& \frac{\left|\Sigma^{-\frac{1}{2}}\right|}{(2\pi)^{\frac{d}{2}}}\exp\left(-\frac{1}{2} (\Sigma^{-\frac{1}{2}}(y-\mu))^\top  \Sigma^{-\frac{1}{2}}(y-\mu)\right),\\
&=& \frac{1}{(2\pi)^{\frac{d}{2}} \sqrt{|\Sigma|}}\exp\left(-\frac{1}{2} (y-\mu)^\top \Sigma^{-1} (y-\mu) \right),\label{eq:mvndensity}
\end{eqnarray}
since $\left|\Sigma^{-\frac{1}{2}}\right|=\frac{1}{\left|\Sigma^{\frac{1}{2}}\right|}=\frac{1}{\sqrt{|\Sigma|}}$ and $(\Sigma^{-\frac{1}{2}}(y-\mu))^\top=(y-\mu)^\top \Sigma^{-\frac{1}{2}}$ since $\Sigma=\Sigma^\top$ (and by using the matrix trace cyclic property).
Eq.~\ref{eq:mvndensity} recovers the multivariate normal density.
It follows that if $X\sim \calN(\mu,\Sigma)$ then we have $Y=\Sigma^{-\frac{1}{2}}(X-\mu)\sim \calN(0,I)$.

Multivariate location-scale families include the {\em elliptical families} which have densities of the form~\cite{StatMatrices-2006}:
\begin{equation}
p_{\mu,V}^\elliptical(x) = |V|^{-\frac{1}{2}} h\left((x-\mu)^\top V^{-1} (x-\mu)\right),
\end{equation}
where $h$ is a profile function.
Indeed, let $P=V^{\frac{1}{2}}$ and $\mu=l$ with $p(x)=h(x^\top x)$.
Then we have
\begin{eqnarray}
p_{\mu,V^{\frac{1}{2}}} &=& |V|^{-\frac{1}{2}} h\left((V^{-\frac{1}{2}}(x-\mu))^\top (V^{-\frac{1}{2}}(x-\mu))\right),\\
 &=& |V|^{-\frac{1}{2}}  h\left((x-\mu)^\top V^{-1} (x-\mu)\right) := p_{\mu,V}^\elliptical(x).
\end{eqnarray} 

Moreover, the elliptical families include  the spherical subfamilies as a special case when $P=I$, see ~\cite{StatMatrices-2006}.
Last, let us remark that some parametric families of distributions can be both interpreted as  location-scale families and exponential families~\cite{EF-2014} (e.g., normal  family, Rayleigh family, inverse Gaussian family, and gamma family).
 
The multivariate location-scale group $\bbG_d$ can be defined on the multivariate location-scale parameter space $G_d=\bbR^d\times\bbP_{++}^d$. 
The identity element is $\mathrm{id}=(0,I)$, the group operation is
$g_{l_2,P_2}.g_{l_1,P_1}=g_{l_2+P_2l_1,P_2P_1}$.
This group operation rule can be found by the action of the location-scale group onto the standard density:
\begin{eqnarray}
g_{l_2,P_2}.g_{l_1,P_1}\odot p(x)&=&|P_2|^{-1} |P_1|^{-1} p\left(P_1^{-1}(P_2^{-1}(x-l_2)-l_1)\right),\\
&=& (P_2P_1)^{-1} p\left( (P_2P_1)^{-1}x -(P_2P_1)^{-1}l_2-P_1^{-1}l_1   \right),\\
&=& (P_2P_1)^{-1} p\left( (P_2P_1)^{-1} (x -l_2-P_2l_1)\right).
\end{eqnarray}

The action of the multivariate location-scale group on a density $p$ is given by:
\begin{equation}
g_{l,P}\odot p := |P|^{-1}\, p\left( |P|^{-1}(x-l) \right).
\end{equation}
The multivariate location-scale family (i.e., set of location-scale models) is obtained by taking the group orbit of the standard density $p$:
\begin{equation}
\calF_p = G_d\odot p.
\end{equation}
Thus the location-scale group $(G_d,.,\mathrm{id})$ is represented by the location-scale matrix group $(\bbG_d,\times,I_{d+1})$.

The corresponding multivariate location-scale block matrix group is the following set of $(d+1)\times (d+1)$ matrices: 
\begin{equation}
\bbG_d=\left\{ M_{l,P}=\mattwotwo{P}{l}{0_d^\top}{1} \st (l,P)\in\bbR^d\times\bbP_{++}^d \right\},
\end{equation}
The inverse element $g^{-1}_{l,P}=g_{-P^{-1}l,P^{-1}}$ can be found from the matrix inverse of $M_{l,P}$.
Indeed, we check that:
\begin{equation}
\mattwotwo{P}{l}{0_d^\top}{1}
\mattwotwo{P^{-1}}{-P^{-1}l}{0_d^\top}{1}
=
\mattwotwo{I}{0_d}{0_d^\top}{1}=I_{d+1}.
\end{equation}
The matrix group multiplication is
\begin{equation}
\mattwotwo{P_1}{l_1}{0_d^\top}{1}\times \mattwotwo{P_2}{l_2}{0_d^\top}{1}
=
\mattwotwo{P_1P_2}{P_1l_2+l_1}{0_d^\top}{1}.
\end{equation}

The Fisher-Rao geometry and  $\alpha$-geometry of multivariate normal distributions was studied in~\cite{Skovgaard-1984} and is reviewed in~\cite{SIMBAD-2013}.
More generally, Mitchell studied the $\alpha$-geometry of elliptical families~\cite{Mitchell-1989}.
Ohara and Eguchi~\cite{Ohara-2014} studied some dually flat geometry of elliptical families.
Warped Riemannian metrics have also been studied for location-scale families defined on a Riemannian manifold~\cite{WarpedRieMetricsLS-2019} (including the Euclidean manifold $\bbR^d$): 
For example, the family of $d$-dimensional isotropic normal distributions is a multivariate location family whose Fisher-Rao metric is a warped Riemannian metric.

\section{Statistical divergences between location-scale densities}\label{sec:fdiv}
Let us consider the statistical $f$-divergences~\cite{Csiszar-1967} $I_f$ between two continuous distributions $p$ and $q$ of $\bbR^d$:
\begin{equation}
I_f(p:q) = \int_{x\in\calX} p(x) f\left(\frac{q(x)}{p(x)}\right) \dx,
\end{equation}
where $f$ is a convex function, strictly convex at $1$, satisfying $f(1)=0$.
When the $f$-divergence generator is chosen to be $f(u)=-\log(u)$, we retrieve the Kullback-Leibler divergence (KLD):
\begin{equation}
D_\KL(p:q)=\int p(x)\log\frac{p(x)}{q(x)}\dx.
\end{equation}
The reverse $f$-divergence $I_f^r(p:q):=I_f(q:p)$ is obtained for the {\em conjugate generator} $f^*(u):=uf\left(\frac{1}{u}\right)$ (convex with $f^*(1)=0$):
$I_f^r(p:q)=I_{f^*}(p:q)=I_f(q:p)$.

Let  $p=p_{0,I}$ and $q=q_{0,I}$ be the two standard PDFs with support $\bbR^d$ defining multivariate location-scale families $\calF_p$ and $\calF_q$, respectively.
Let $p_{l_1,P_1}\in\calF_p$ and $q_{l_2,P_2}\in\calF_q$.

We state the following group invariance theorem of the $f$-divergences:

\begin{BoxedTheorem}[Invariance of $f$-divergences under the location-scale group]\label{thm:invgroupfdiv}
We have 
$$
I_f(g\odot p:g\odot q)=I_f(p:q)
$$ 
for all $p,q\in\calP(\bbR^d)$ and any $g=g_{l,P}$ in the multivariate location-scale group $G_d=\bbR^d\times\bbP_{++}^d$.
\end{BoxedTheorem}

\begin{proof}
We have
\begin{eqnarray}
I_f(g\odot p:g\odot q) &=& \int |P|^{-1} p\left(|P|^{-1}(x-l)\right) \log 
\left(\frac{|P|^{-1} p\left(|P|^{-1}(x-l)\right)}{|P|^{-1} q\left(|P|^{-1}(x-l)\right)}\right) \dx,\\
&=& \int p\left(|P|^{-1}(x-l)\right) \log \left(\frac{p\left(y\right)}{q\left(y\right)}\right) \dy =: I_f(p:q),
\end{eqnarray}
after making a change of variable $y=|P|^{-1}(x-l)$ in the multiple integral   $\int_\calX \ldots \dx=\int_{\bbR}\ldots \int_{\bbR} \ldots \dx_1\ldots\dx_d$ with $\dy=|P|^{-1}\dx$.
This change of variable {\em requires} $\calX=\bbR^d$~\cite{Lax-2001} and therefore $p,q\in\calP(\bbR^d)$.
Indeed, when the support of the PDFs are dependent of $(l,P)$ (e.g., a uniform distribution on a compact $\calK\subset\bbR^d$), the KLD   diverges and the Fisher information is infinite~\cite{Hayashi-2011}. 
Thus we assume in the remainder that all location-scale families are regular.
\end{proof}

From Theorem~\ref{thm:invgroupfdiv}, we get the following corollary:

\begin{BoxedCorollary}[Canonical settings for $f$-divergences between location-scale distributions]\label{thm:reducedfdiv}
The $f$-divergence between two regular location-scale densities is equivalent to the $f$-divergence between one standard location-scale density and another affinely shifted location-scale density:
\begin{equation}\label{eq:fdivmvls}
I_f(p_{l_1,P_1}:q_{l_2,P_2}) = I_f\left( p : q_{ P_1^{-1}(l_2-l_1), P_1^{-1}P_2}\right)
=I_f\left(p_{P_2^{-1}(l_1-l_2),P_2^{-1}P_1}:q\right).
\end{equation}
\end{BoxedCorollary}

\begin{proof}

We give two proofs: A short indirect proof relying on Theorem~\ref{thm:invgroupfdiv} and a direct proof.

\begin{itemize}
\item
Let $g_1=g_{l_1,P_1}$ and $g_2=g_{l_2,P_2}$ so that $p_{l_1,P_1}=p_{g_1}$ and $q_{l_2,P_2}=q_{g_2}$. 
Applying Theorem~\ref{thm:invgroupfdiv} with $g=g_1$, we 
have $I_f(g_1\odot p: g_2\odot q)=I_f(g_1^{-1}.g_1\odot p: g_1^{-1}.g_2\odot q)$.
Since $g_1^{-1}.g_1=\mathrm{id}$ and $g_1^{-1}.g_2=g_{P_1^{-1}(l_2-l_1), P_1^{-1}P_2}$, we get 
$I_f(p_{l_1,P_1}:q_{l_2,P_2}) =I_f\left( p : q_{ P_1^{-1}(l_2-l_1), P_1^{-1}P_2}\right)$.
Similarly, Applying Theorem~\ref{thm:invgroupfdiv} with $g=g_2$, we get $I_f(p_{l_1,P_1}:q_{l_2,P_2})=I_f\left(p_{P_2^{-1}(l_1-l_2),P_2^{-1}P_1}:q\right)$
since $g_2^{-1}.g_1=g_{P_2^{-1}(l_1-l_2),P_2^{-1}P_1}$.

\item The second direct proof makes the change of variable in $\stackrel{\star}{=}$ with
 $y=P_1^{-1}(x-l_1)$, $x=P_1y+l_1$, $\dy=|P_1|^{-1} \dx$ and $\dx=  |P_1| \dy$, and uses the identity $\frac{|P_2|^{-1}}{|P_1|^{-1}}=|P_1^{-1}P_2|^{-1}$:
\begin{eqnarray}
I_f(p_{l_1,P_1}:q_{l_2,P_2}) &:=& 
\int_{\calX} p_{l_1,P_1}(x)\, f\left(\frac{q_{l_2,P_2}(x)}{p_{l_1,P_1}(x)}\right) \dx,\\
&=& \int |P_1|^{-1} \,  p\left(P_1^{-1}(x-l_1)\right)\,
 f\left( \frac{|P_2|^{-1} \, q\left( P_2^{-1}(x-l_2)\right)}{|P_1|^{-1}\,  p\left( P_1^{-1} (x-l_1)\right)}\right) \dx,\nonumber\\
&\stackrel{\star}{=}& 
\int p(y)\, f\left( \frac{|P_2|^{-1}}{|P_1|^{-1}}\,  \frac{q(P_2^{-1}(P_1y+\mu_1)-\mu_2))}{  p(y)}\right) \dy ,\nonumber\\
&=& 
\int p(y)\, f\left( |P_1^{-1}P_2|^{-1} \frac{q((P_1^{-1}P_2)^{-1}(y-P_2^{-1}(l_2-l_1)))}{  p(y)}\right) \dy ,\\
&=& I_f\left( p : q_{ P_2^{-1}(l_2-l_1),P_2P_1^{-1}}\right),
\end{eqnarray}
Using the conjugate generator $f^*(u)$, we get $I_f(p_{l_1,P_1}:q_{l_2,P_2}) = I_f\left(p_{P_2^{-1}(l_1-l_2),P_2^{-1}P_1}:q\right)$.
\end{itemize}
\end{proof}

Thus we obtain the scale invariance of the $f$-divergence between multivariate scale families (including zero-centered elliptical distributions):

\begin{BoxedCorollary}[Scale invariance of $f$-divergences between scale densities]\label{thm:fdivscaleinvariance}
The $f$-divergence between  multivariate scale densities $p_{P_1}$ and $q_{P_2}$ is scale-invariant:
For all $\lambda>0$: $I_f({p_{\lambda P_1}:p_{\lambda P_2}})=I_f({p_{P_1}:p_{P_2}})=I_f(p:q_{P_1^{-1}P_2})=I_f(p_{P_2^{-1}P_1}:q)$.
\end{BoxedCorollary}

\begin{example}
Consider the Rayleigh scale family with $\calX=\bbR_{++}$ and standard density $p(x)=x\exp(-\frac{x^2}{2})$.
The KLD between two Rayleigh distributions is 
\begin{equation}
D_\KL(p_{\sigma_1^2}:p_{\sigma_2^2})=\frac{\sigma_1^2}{\sigma_2^2}-\log\left(\frac{\sigma_1^2}{\sigma_2^2}\right)-1.
\end{equation} 
We check that $D_\KL(g_{\lambda}\odot p_{\sigma_1^2}:g_{\lambda}\odot p_{\sigma_2^2})=D_\KL(p_{\sigma_1^2}:p_{\sigma_2^2})$ since $g_{\lambda}\odot p_{\sigma^2}=p_{\lambda\sigma^2}$ and 
$\frac{\lambda\sigma_1^2}{\lambda\sigma_2^2}=\frac{\sigma_1^2}{\sigma_2^2}$.
Similarly, the KLD between two univariate  zero-centered normal distributions yields the same formula.
In fact the Rayleigh distributions form an exponential family and the KLD amounts to a Bregman divergence which is the Itakura-Saito divergence $D_\IS(\theta_1:\theta_2):=\frac{\theta_1}{\theta_2}-\log \frac{\theta_1}{\theta_2}-1$.
We have~$D_\KL(p_{\sigma_1^2}:p_{\sigma_2^2})=D_\IS(\theta_2:\theta_1)$ with $\theta_i=-\frac{1}{2\sigma_i^2}$.
See~\cite{EF-2009} for details.
\end{example}

Let us instantiate the invariance property of Corollary~\ref{thm:reducedfdiv} for the KLD. 
We get:

\begin{BoxedCorollary}[KLD between location-scale densities]\label{thm:kldls}
The KLD  between two regular location-scale densities is equivalent to the $f$-divergence between one standard location-scale density and another affinely shifted location-scale density:
\begin{equation}\label{eq:kldmvls}
D_\KL(p_{l_1,P_1}:q_{l_2,P_2}) = D_\KL\left( p : q_{ P_1^{-1}(l_2-l_1), P_1^{-1}P_2}\right)
=D_\KL\left(p_{P_2^{-1}(l_1-l_2),P_2^{-1}P_1}:q\right).
\end{equation}
\end{BoxedCorollary}

Since KLD $D_\KL(p:q)$ amounts to the cross-entropy $h^\times(p:q)=-\int p(x)\log q(x)\dx$ minus Shannon's differential entropy $h(p)=h^\times(p:p)=-\int p(x)\log p(x)\dx$, let us also report the formula for the cross-entropy/entropy under the action of a location-scale group element $g=g_{l,P}$:

\begin{eqnarray}
h^\times(g\odot p: g\odot q)  &=& h^\times(p:q) + \log |P|,\\
h(g\odot p) &=& h(p)+ \log |P|.
\end{eqnarray}
Thus $D_\KL(g\odot p: g \odot q)=h^\times(g\odot p: g\odot q) -h(g\odot p)=h^\times(p:q)- h(p)=D_\KL(p:q)$.

Furthermore, we have:
\begin{eqnarray}
h^\times(p_{l_1,P_1}: q_{l_2,P_2})  &=& h^\times(p:q_{ P_1^{-1}(l_2-l_1), P_1^{-1}P_2}) - \log |P_1|,\\
 &=& h^\times(p_{P_2^{-1}(l_1-l_2),P_2^{-1}P_1}:q) - \log |P_2|.
\end{eqnarray}

Notice that it is well-known that the $f$-divergence between two continuous densities with full support in $\bbR^d$ is independent of a diffeomorphism~\cite{fdivdiffeo-2010} $Y=t(X)$: That is, 
$I_f(p_X(x):q_X(x))=I_f(p_Y(y):q_Y(y))$. 
The proof also makes use of a change of variable in a multiple integral and requires~\cite{Lax-2001} $\calX=\bbR^d$:

\begin{BoxedProposition}[Invariance of $f$-divergences]\label{prop:fdivinvariance}
Let $t:\bbR^d\rightarrow\bbR^d$ be a diffeomorphism, $p_X, q_X\in\calP(\bbR^d)$ and $Y=t(X)$. 
Then we have $I_f(p_Y(y):q_Y(y))=I_f(p_X(x):q_X(x))$.
\end{BoxedProposition}

\begin{proof}
Let $p_Y(y)= |\Jac_{t^{-1}}(y)| \times p_X(t^{-1}(y))$ and $q_Y(y)= |\Jac_{t^{-1}}(y)| \times q_X(t^{-1}(y))$ with $x=t^{-1}(y)$ and 
$\dx=|\Jac_{t^{-1}}(y)|\dy$. We have:
\begin{eqnarray}
I_f(p_Y:q_Y) &=&  \int_{\bbR^d} p_Y(y) f\left( \frac{q_Y(y)}{p_Y(y)}\right) \dx \\
&=& \int_{\bbR^d} |\Jac_{t^{-1}}(y)| \times p_X(t^{-1}(y)) f\left( \frac{|\Jac_{t^{-1}}(y)| \times q_X(t^{-1}(y))}{|\Jac_{t^{-1}}(y)| \times p_X(t^{-1}(y))} \right) \dy,\\
&=& \int_{\bbR^d} p_X(x)f\left( \frac{q_X(x)}{p_X(x)}\right) \dx =: I_f(p_X:q_X).
\end{eqnarray}
\end{proof}
Letting $Y=PX+l$, $p_Y=g_{l,P}\odot p_X$ and $q_Y=g_{l,P}\odot q_X$, we get $I_f(g_{l,P}\odot p_X :g_{l,P}\odot q_X)=I_f(p_X : q_X)$.

\begin{example}\label{ex:lognormal}
Consider the family of log-normal distributions~\cite{LogNormal-1987} such that if $X\sim\calN(\mu,\sigma)$ then $Y=\exp(X)$ follows a log-normal distribution $\calLN(\mu,\sigma)$ with probability density function:
\begin{equation}
p^\calLN_{\mu,\sigma}(x):=\frac{1}{x \sigma \sqrt{2 \pi}} \exp \left(-\frac{(\ln x-\mu)^{2}}{2 \sigma^{2}}\right),
\end{equation}
for $x\in\calX=(0,\infty)$.
Reciprocally, if $X\sim\calLN(\mu,\sigma)$ then $Y=\log(X)$ follows a normal distribution $\calN(\mu,\sigma)$.
It follows from Proposition~\ref{prop:fdivinvariance} that the $f$-divergence $I_f(p^\calLN_{\mu_1,\sigma_1}:p^\calLN_{\mu_2,\sigma_2})=I_f(p^\calN_{\mu_1,\sigma_1}:p^\calN_{\mu_2,\sigma_2})$ (see also~\cite{Renyi-2013} for the matching formula of the Kullback-Leibler divergence).
\end{example} 

We can define the {\em $f$-mutual information} between two random variables $X$ and $Y$ as 
\begin{equation}
\MI_f(X;Y):=I_f(p_{(X,Y)}:p_Xp_Y).
\end{equation}
Whenever $p_{(X,Y)}=p_Xp_Y$, we say that random variable $X$ is independent to random variable $Y$, and the $f$-mutual information is zero: $\MI_f(X;Y)=0$. 
We have the following invariance of the mutual information:

\begin{BoxedProposition}[Invariance of $f$-mutual information] 
For any invertible and differentiable transformations $t_1$ and $t_2$ from $\bbR^d$ to $\bbR^d$, we have
$\MI_f(t_1(X_1);t_2(X_2))=\MI_f(X_1:X_2)$.
\end{BoxedProposition}

\begin{proof}
Let $Y_1=t_1(X_1)$ and $Y_2=t_2(X_2)$.
We have the joint density $p_{(Y_1,Y_2)}(y_1,y_2)=|\Jac_{t_1^{-1}}(y_1)|\ |\Jac_{t_2^{-1}}(y_2)|\  p_{(X_1,X_2)}(x_1,x_2)$
and the marginals $p_{Y_1}(y_1)=|\Jac_{t_1^{-1}}(y_1)|\  p_{X_1}(x_1)$ and $p_{Y_2}(y_2)=|\Jac_{t_2^{-1}}(y_2)|\  p_{X_2}(x_2)$.
It follows that $\frac{p_{Y_1}(y_1)p_{Y_2}(y_2)}{p_{(Y_1,Y_2)}(y_1,y_2)}=\frac{p_{X_1}(x_1)p_{X_2}(x_2)}{p_{(X_1,X_2)}(x_1,x_2)}$.
The $f$-mutual information $\MI_f(t_1(X_1);t_2(X_2))$ rewrites as:
\begin{eqnarray}
\MI_f(t_1(X_1);t_2(X_2)) &=& \int_{y_1}\int_{y_2} p_{(Y_1,Y_2)}(y_1,y_2) f\left(\frac{p_{Y_1}(y_1)p_{Y_2}(y_2)}{ p_{(Y_1,Y_2)}(y_1,y_2)} \right)  \dy_1\dy_2,\\
&=& \int_{y_1}\int_{y_2} p_{(Y_1,Y_2)}(y_1,y_2)  f\left(\frac{p_{X_1}(x_1)p_{X_2}(x_2)}{ p_{(X_1,X_2)}(x_1,x_2)} \right)   \dy_1\dy_2.
\label{eq:miproofint}
\end{eqnarray}
Using two changes of variables $x_1=t_1^{-1}(y_1)$ and $x_2=t_2^{-1}(x_2)$
with $|\Jac_{t_1^{-1}}(y_1)|\ \dy_1=\dx_1$ and $|\Jac_{t_2^{-1}}(y_2)|\ \dy_2=\dx_2$, we have:
\begin{eqnarray}
p_{(Y_1,Y_2)}(y_1,y_2)\dy_1\dy_2 &=& |\Jac_{t_1^{-1}}(y_1)|\ |\Jac_{t_2^{-1}}(y_2)|\  p_{(X_1,X_2)}(x_1,x_2)  \dy_1\dy_2,\\
&=&  p_{(X_1,X_2)}(x_1,x_2)  \dx_1\dx_2.
\end{eqnarray}
Thus we have Eq.~\ref{eq:miproofint} which rewrites as:
\begin{eqnarray}
\MI_f(t_1(X_1);t_2(X_2)) &=&\int_{x_1}\int_{x_2} p_{(x_1,x_2)}(x_1,x_2)  f\left(\frac{p_{X_1}(x_1)p_{X_2}(x_2)}{ p_{(X_1,X_2)}(x_1,x_2)} \right) \dx_1\dx_2,\\
&=:& \MI_f(X_1:X_2).
\end{eqnarray}
Notice that for the change of variables we require to have both the joint densities and the marginal densities to be defined on the full support $\bbR^d$~\cite{Lax-2001}. 
\end{proof}

Let us illustrate the formula of Eq.~\ref{eq:kldmvls} in the following example:

\begin{example}
The KLD between the standard normal   $p^\normal$ and a  normal $p_{\mu,\sigma}^\normal=p_{\mu,\sigma}$ is
\begin{equation}
D_\KL\left(p^\normal:p_{\mu,\sigma}^\normal\right)= \frac{\mu^2}{2\sigma^2}+\frac{1}{2}\left(\frac{1}{\sigma^2}-\log\frac{1}{\sigma^2}-1\right).
\end{equation}

From this formula, we recover the generic KLD formula between two normal distributions by plugging $\sigma=\frac{\sigma_2}{\sigma_1}$
 and $\mu=\frac{\mu_2-\mu_1}{\sigma_1}$:
\begin{eqnarray}
D_\KL\left(p_{\mu_1,\sigma_1}^\normal:p_{\mu_2,\sigma_2}^\normal\right)&=& D_\KL\left(p^\normal:p_{\frac{\mu_2-\mu_1}{\sigma_1},\frac{\sigma_2}{\sigma_1}}^\normal\right),\\
&=& \frac{(\mu_2-\mu_1)^2}{2\sigma_2^2}+\frac{1}{2}\left(\frac{\sigma_1^2}{\sigma_2^2}-\log\frac{\sigma_1^2}{\sigma_2^2} -1\right).
\end{eqnarray}

Equivalently, we could also have used the canonical formula:
\begin{equation}
D_\KL\left(p_{\mu_2,\sigma_2}^\normal:p^\normal\right)=\frac{1}{2}\left(\sigma^2+\mu^2-1-\log\sigma^2\right),
\end{equation}
 and then retrieve the ordinary formula as follows:
\begin{eqnarray}
D_\KL\left(p^\normal:p_{\mu,\sigma}^\normal\right)&=&D_\KL\left(p^\normal_{\frac{\mu_1-\mu_2}{\sigma^2},\frac{\sigma_1}{\sigma_2}}:p^\normal\right),\\
&=& \frac{(\mu_2-\mu_1)^2}{2\sigma_2^2}+\frac{1}{2}\left(\frac{\sigma_1^2}{\sigma_2^2}-\log\frac{\sigma_1^2}{\sigma_2^2} -1\right).
\end{eqnarray}

The KLD between the standard multivariate normal (MVN) $p^\normal$ and a multivariate normal $p_{\mu,\Sigma}^\normal=p_{\mu,\Sigma^{\frac{1}{2}}}$ is
\begin{equation}
D_\KL\left(p^\normal:p_{\mu,\Sigma}^\normal\right)=\frac{1}{2}\left( \tr(\Sigma^{-1})+\mu^\top\Sigma^{-1}\mu+\log |\Sigma|-d\right).
\end{equation}

Using Corollary~\ref{thm:kldls}, we recover the formula for the KLD between two normal distributions with $\Sigma=\Sigma_1^{-1}\Sigma^2$ and $\mu=\Sigma_1^{-\frac{1}{2}} (\mu_2-\mu_1)$:
\begin{eqnarray}
D_\KL\left(p_{\mu_1,\Sigma_1}^\normal:p_{\mu_2,\Sigma_2}^\normal\right)
&=& D_\KL\left(p:p_{\Sigma_1^{-\frac{1}{2}}(\mu_2-\mu_1),\Sigma_1^{-1}\Sigma_2}\right),\\
&=& \frac{1}{2}\left( \tr(\Sigma_2^{-1}\Sigma_1)+
(\mu_2-\mu_1)^\top \Sigma_2^{-1}(\mu_2-\mu_1)+\log |\Sigma_1^{-1}\Sigma_2|-d\right).
\end{eqnarray}
Observe that the KLD between two multivariate normal distributions can be decomposed as the sum of a squared Mahalanobis distance 
\begin{equation}
D_\Mah^{Q}(\mu_1,\mu_2):=\frac{1}{2}(\mu_2-\mu_1)^\top Q (\mu_2-\mu_1),
\end{equation} for $Q\succ 0$,  and a scale-invariant matrix Itakura-Saito divergence
 \begin{equation}
D_\IS(\Sigma_1,\Sigma_2):=\frac{1}{2} \left( \tr(\Sigma_2^{-1}\Sigma_1-I) - \log |\Sigma_2^{-1}\Sigma_1| \right),
\end{equation}
 also called Burg matrix divergence in~\cite{MVNClustering-2007}, a matrix Bregman divergence~\cite{NN-2013}):
\begin{equation}
D_\KL\left(p_{\mu_1,\Sigma_1}^\normal:p_{\mu_2,\Sigma_2}^\normal\right)
= D_\Mah^{\Sigma_2^{-1}}(\mu_1,\mu_2) + D_\IS(\Sigma_1,\Sigma_2).
\end{equation}

\end{example}

We can also derive similar results for the {\em linear group} $Y=AX+b$ of transformations for $A\in\GL(d)$ (group of invertible $d\times d$ matrices) and $b\in\bbR^d$.


\section{Monte Carlo estimators of $f$-divergences}\label{sec:MCfdiv}

Depending on the standard densities $p$ and $q$, the integrals of the $f$-divergences may be calculable in closed-form or not.
When no closed-form is available, we can {\em estimate} the $f$-divergences using Monte Carlo importance sampling~\cite{MC-2013} as follows:
We choose a {\em propositional distribution} $r$ and use a set $\calS_m=\{x_1,\ldots, x_m\}\sim_{\mathrm{iid}} r$ of $m$ i.i.d. variates sampled from $r$ to estimate the $f$-divergence as follows:
\begin{equation}
\hat{I}_{f,\calS_m}(p:q) =   \frac{1}{m} \sum_{i=1}^m \frac{p(x_i)}{r(x_i)} f\left(\frac{q(x_i)}{p(x_i)}\right).
\end{equation}
In particular, when $r=p$, we end up with the following estimate often met in the literature:
\begin{equation}
\hat{I}_{f,\calS_m}(p:q) =   \frac{1}{m} \sum_{i=1}^m  f\left(\frac{q(x_i)}{p(x_i)}\right).  
\end{equation}
For example, we estimate the Kullback-Leibler divergence by 
$\hat{D}_{\KL,\calS_m}(p:q) =   \frac{1}{m} \sum_{i=1}^m  \log\left(\frac{p(x_i)}{q(x_i)}\right)$.

One of the problem of MC estimators is that they may yield inconsistent divergence measures when the proposal distribution depends on the arguments of the $f$-divergences. That is one realization (i.e., sampling with $\calS_m$) may find that $\hat{I}_{f,\calS_m}(p_1:q)> \hat{I}_{f,\calS_m}(p_2:q)$ while another
realization (i.e., sampling with $\calS_m'$) may find that opposite result $\hat{I}_{f,\calS_m'}(p_1:q)< \hat{I}_{f,\calS_m'}(p_2:q)$.
This lack of consistency is problematic when implementing algorithms based on divergence comparison predicates.

However, since for location-scale densities we can always reduce the calculation of $f$-divergences using one standard density, say:
\begin{equation}
{I}_{f}(p_{l_1,P_1}:q_{l_2,P_2}) =    {I}_{f}(p:q_{P_1^{-1}(l_2-l_1),P_1^{-1}P_2}),
\end{equation}
we can  estimate the $f$-divergences with a fixed set $\calS_m$ of iid. random variates sampled from the standard density $p$ as follows:
\begin{eqnarray}
\hat{I}_{f,\calS_m}(p_{l_1,P_1}:q_{l_2,P_2}) &=&   
 \hat{I}_{f,\calS_m}(p:q_{P_1^{-1}(l_2-l_1),P_1^{-1}P_2}),\\
&=& \frac{1}{m} \sum_{i=1}^m  f\left(\frac{q_{P_1^{-1}(l_2-l_1),P_1^{-1}P_2}(x_i)}{p(x_i)}\right).
\end{eqnarray}

Another problem when  estimating the $f$-divergences with Monte Carlo methods is that depending on the randomly sampled variates, we may 
 end up with negative estimates.
To overcome this problem, we shall use the following identity:
\begin{eqnarray}
I_f(p:q) &=& \int p(x) B_f\left(\frac{q(x)}{p(x)}:1\right) \dx = E_p\left[B_f\left(\frac{q(x)}{p(x)}:1\right)\right],
\end{eqnarray}
where  where $B_f(a:b)$
is the {\em scalar Bregman divergence}~\cite{BD-1967}:
\begin{equation}\label{eq:bds}
B_f(a:b)= f(a)-f(b)-(a-b)f'(b)\geq 0.
\end{equation}
Indeed, since $f(1)=0$, we have
\begin{eqnarray}
\int p(x) B_f\left(\frac{q(x)}{p(x)}:1\right)\dx
 &=& \int p(x) \left( f\left(\frac{q(x)}{p(x)}\right) - \left(\frac{q(x)}{p(x)}-1\right)f'(1) \right)\dx,\\
&=& \int p(x) f\left(\frac{q(x)}{p(x)}\right)\dx - f'(1) \underbrace{\int (q(x)-p(x))\dx}_{0}
 =: I_f(p:q). 
\end{eqnarray}

Since Bregman divergences are always non-negative and equal to zero iff $a=b$, we get another proof of Gibbs' inequality $I_f(p:q)\geq 0$ (complementing the proof using Jensen's inequality).
Thus we can estimate the $f$-divergences non-negatively using iid. random variates $x_1,\ldots, x_m$ from $p(x)$ as follows:
\begin{equation}
\hat{I}_f(p_{l_1,P_1}:q_{l_2,P_2}) = \frac{1}{m}\sum_{i=1}^m B_f\left(\frac{q_{P_1^{-1}(l_2-l_1),P_1^{-1}P_2}(x_i)}{p(x_i)}:1\right)\geq 0.
\end{equation}

Furthermore, since the MC estimator of the $f$-divergence is the average of $m$ scalar Bregman divergences, it follows that the estimator is a {\em proper divergence} (i.e, $\hat{I}_f(p_{l_1,P_1}:p_{l_2,P_2})=0 \Leftrightarrow (l_1,P_1)=(l_2,P_2)$) whenever two distinct densities of the location-scale families cannot coincide in more than $s$ points and when the random variates $x_i$'s have at least $s+1$ distinct points.

\section{Information projections onto location-scale families}\label{sec:infproj}

We investigate how any two location-scale models $\calF_p$ and $\calF_q$ (with $p\not=q$ and $p,q\in\calP(\calX)$) 
relate to each other using information projections induced by $f$-divergences~\cite{InfoProj-1975,InformationProjection-2018}.
For a family of densities $\calQ$, let $I_f(p:\calQ):=\inf_{q\in\calQ} I_f({p:q})$ (respectively, 
$I_f(\calP:q):=\inf_{p\in\calP} I_f({p:q})$).
We consider the (possibly multivariate) location-scale models as subspaces of $\calP(\bbR^d)$ (infinite-dimensional space) or as submodels of a multivariate location-scale model $\calF_m$. 
In the former case, we may consider nonparametric information geometry~\cite{DensityMfd-1988,NIG-1995,NIG-2013} for geometrically modeling $\calP(\calX)$.
In the latter case, we consider the ordinary statistical manifold structure of $\calF_m$ (parametric information geometry~\cite{IG-2016,EIG-2020}).
First, let us observe that even if the KLD is asymmetric, one orientation can be finite while the reverse orientation can be infinite:
For example, we have $D_\KL(p^\calN:p^\calC)\simeq 0.26<\infty$ but $D_\KL(p^\calC:p^\calN)=+\infty$ where $p^\calN$ denotes the standard normal density and $p^\calC$ denotes the standard Cauchy density (heavy-tailed).

Recall that $G_d=\bbR^d\times\mathbb{P}_{++}^d$ denotes the $d$-dimensional location-scale group (or ``location-positive'' group).
We state the remarkable projection property of a location-scale density onto another location-scale model:

\begin{BoxedTheorem}[Information projection on location-scale families]\label{thm:projls}
The $f$-divergence $I_f(p_g:q_{h^*})=I_f(p_g:\calF_q)$ induced by the right-sided $f$-divergence minimization of $p_g\in G_d$ with 
$\calF_q$  is independent of $g$, 
i.e. $I_f(p_g:\calF_q)=I_f(p_{g'}:\calF_q)$ for all $g'\in G_d$.
Similarly, the $f$-divergence $I_f(p_{g^*}:q_{h})=I_f(\calF_p:q_h)$ induced by the left-sided  $f$-divergence minimization of $q_h$ with $\calF_p$ is independent of $h$, i.e. $I_f(\calF_p:q_h)=I_f(\calF_p:q_{h'})$ for all $h'\in G_d$.
\end{BoxedTheorem}

\begin{proof} 
Using the invariance of the $f$-divergence under the action of $g^{-1}$ (Theorem~\ref{thm:invgroupfdiv}), we have
\begin{eqnarray}
\inf_{h\in G_d} I_f(p_g:q_h) &=& \inf_{h\in G_d} I_f(g^{-1}\odot p_g:g^{-1}\odot q_h),\\
&=& \inf_{\{h'=g^{-1}.h \ : \ h\in G_d \in G_d\}} I_f(p: q_{h'}),\\
&=& \inf_{h'\in G_d} I_f(p: q_{h'}),
\end{eqnarray}
since the left coset $g^{-1}.G_d$ is equal to $G_d$: 
Indeed, for any $e\in G_d$, we may find $f\in G_d$ such that $g^{-1}.f=e$ (i.e., choose $f=g.e$).
Let $h^*\in G_d$ such that $\inf_{h\in G_d} I_f(p:q_h)=I_f(p:q_{h^*})$.
Thus a minimum of $\inf_{h\in G_d} I_f(p_g:q_h)$ is $h^*(g):=g.h^*$ since 
\begin{equation}
\inf_{h\in G_d} I_f(p_g:q_h) = I_f(p:q_{h^*}) = I_f(p_g:q_{g.{h^*}})=I_f(p_g:q_{h^*(g)}).
\end{equation}

Similarly, by using the conjugate generator $f^*$, we prove that $I_f(\calF_p:q_h)$ is independent of $h$, and  
a minimum of $\inf_{g\in G_d} I_f(p_g:q_h) $ is $g^*(h):=h.g^*$ since 
\begin{equation}
\inf_{g\in G_d} I_f(p_g:q_h) = I_f(p_{g^*}:q) = I_f(p_{h.g^*}:q_{h})=I_f(p_{g^*(h)}:q_{h}).
\end{equation}
\end{proof}

This property was observed without any proof in~\cite{Villa-2016} for the special case of the Kullback-Leibler divergence between any two univariate location-scale families. We extended this property with a proof to $f$-divergences between multivariate location-scale families.
Notice that the projections with respect to $f$-divergences link orbits between the subspaces $\calF_p$ and $\calF_q$:
Namely, we have the mappings $g\mapsto h^*(g)=g.h^*$ and $h\mapsto g^*(h):=h.g^*$.

We shall illustrate the theorem on several examples and provide some geometric interpretations of how the location-scale submodels relate to each others.

\begin{figure}
\centering

\includegraphics[width=\textwidth]{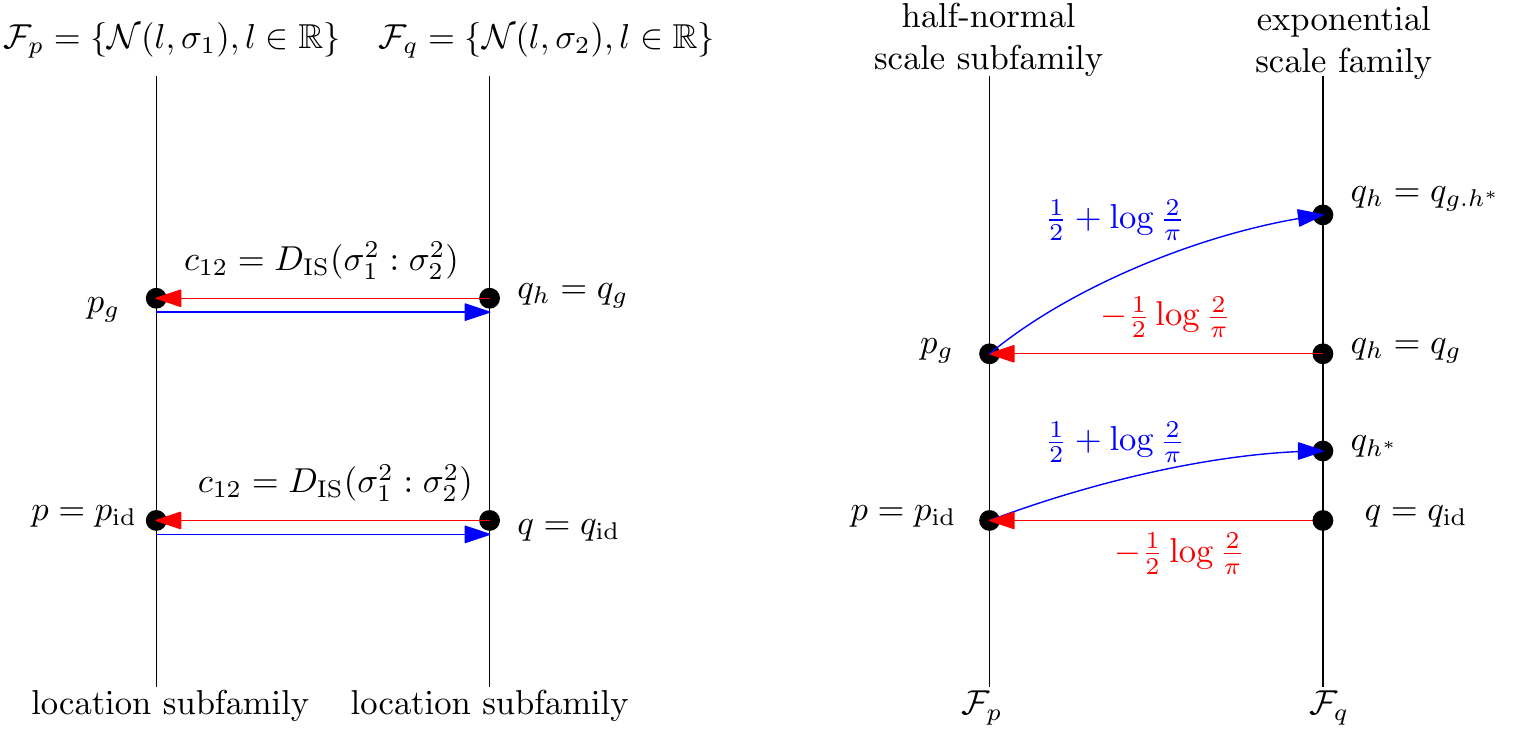}

\caption{Illustrations of the information projections between two location-scale families $\calF_p$ and $\calF_q$.}\label{fig:infproj}
\end{figure}

\begin{example}
The first example consider two location subfamilies of the Gaussian location-scale family: 
Let $p(x):=p^\calN_{l,\sigma_1}(x)$ and 
$q(x):=p^\calN_{l,\sigma_2}(x)$ for prescribed distinct values $\sigma_1\not=\sigma_2$. 
Consider the KLD between one density $p_g$ of $\calF_p$ and another density $q_h$ of $\calF_q$:
\begin{equation}
D_\KL(p_g:q_h)=\frac{(g-h)^2}{2\sigma_2^2}+c_{12},
\end{equation}
where $c_{12}=D_\IS(\sigma_1^2:\sigma_2^2)=\frac{1}{2}\left(\frac{\sigma_1^2}{\sigma_2^2}-\log \frac{\sigma_1^2}{\sigma_2^2}-1\right)$ is a constant.
In that case $D_\KL(p_g:\calF_q)=c_{12}$ and $h^*=\mathrm{id}$ so that $h^*(g)=g.\mathrm{id}=g$, 
and $D_\KL(\calF_p:q_h)=c_{12}$ and $g^*=\mathrm{id}$ so that $g^*(h)=h.g^*=h$.
We may interpret the two location families $\calF_p$ and $\calF_q$ as one-dimensional submanifolds of the dually flat manifold of the family of univariate normal distributions. Then the two submanifolds are at equidivergence from each others as depicted in Figure~\ref{fig:infproj} (left).
\end{example}

The second example reworks the example originally reported in~\cite{Villa-2016}:

\begin{example}
Consider $p(x)=\sqrt{\frac{2}{\pi}}\exp(-\frac{x^2}{2})$ and $q(x)=\exp(-x)$ be the standard density of the half-normal distribution  and the standard density of the exponential distribution defined over the support $\calX=[0,\infty)$, respectively. 
We consider the scale families $\calF_p=\{p_{s_1}(x)=\frac{1}{s_1}p(\frac{x}{s_1}) \st s_1>0\}$ and $\calF_q=\{q_{s_2}(x)=\frac{1}{s_2}q(\frac{x}{s_2}) \st s_2>0\}$.
Using a computer algebra system, we find that
\begin{equation}\label{eq:hne}
D_\KL(p_{s_1}:q_{s_2}) = \frac{1}{2}\left(2\log\frac{s_2}{s_1} + \log\frac{2}{\pi} -1 \right) + \sqrt{\frac{2}{\pi}} \frac{s_1}{s_2}.
\end{equation}

Let $r=\frac{s_1}{s_2}$. Then $D_\KL(p_{s_1}:q_{s_2})=\sqrt{\frac{2}{\pi}}r-\log r+\log \sqrt{\frac{2}{\pi}}-\frac{1}{2}$.
That is, the KLD between the scale families depends only on the scale ratio as proved in Corollary~\ref{thm:fdivscaleinvariance}.

The KLD is minimized wrt. to $s_2$ when $-\frac{1}{r}+\sqrt{\frac{2}{\pi}}=0$: That is, when $r=\sqrt{\frac{\pi}{2}}$ (i.e., $s_2=s_1\sqrt{\frac{\pi}{2}}$).
We check that $D_\KL(p_{s_1}:\calF_q)=\frac{1}{2}+\log \frac{2}{\pi}\simeq 0.048$ is independent of $s_1$.
Thus we have $h^*=g_1^*=\sqrt{\frac{\pi}{2}}$ and $g_{s_1}^*=s_1\sqrt{\frac{\pi}{2}}$.

Similarly, we find that $D_\KL(\calF_p:q_{s_2})$ is minimized wrt $s_1$ for $s_1=s_2$.
and we have $D_\KL(\calF_p:q_{s_2})=-\frac{1}{2}\log\frac{2}{\pi}\simeq 0.226$.
Figure~\ref{fig:infproj} (right) illustrates geometrically the information projections between these two scale families. 
\end{example}

\begin{figure}
\centering

\includegraphics[width=0.35\textwidth]{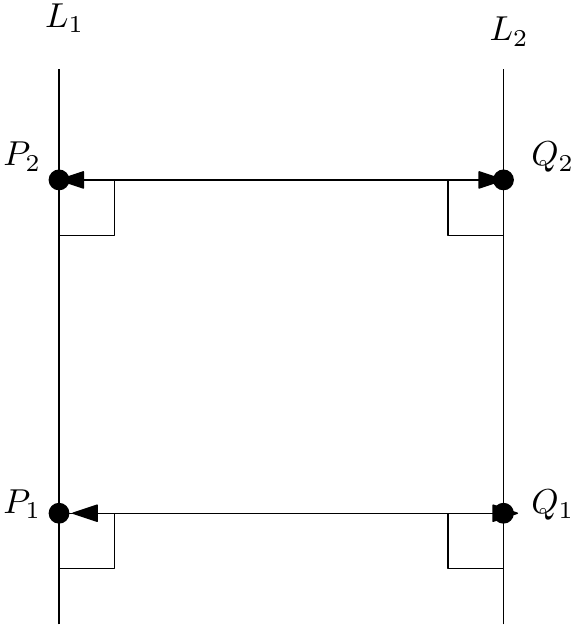}

\caption{In Euclidean geometry, parallel lines $L_1$ and $L_2$ are equidistant to each others.}\label{fig:euclproj}
\end{figure}

Thus the location-scale densities bear some geometric similarity with parallel lines in Euclidean geometry which are equidistant as depicted in Figure~\ref{fig:euclproj}. 

\begin{example}
The Weibull distributions form a one-parametric family of scale families with densities expressed by:
\begin{equation}
p_{k,s}(x)=\frac{k}{s}\left(\frac{x}{s}\right)^{k-1} \exp \left(-\left(\frac{x}{s}\right)^{k}\right),
\end{equation}
for $x\in\calX= [0,\infty)$. Parameter $s$ denotes the scale for location $l=0$.
Let $p_k(x)=p_{k,1}(x)=kx^{k-1}\exp(-x^k)$ denotes the standard density of the $k$-th Weibull scale family.

The Kullback-Leibler divergence between two Weibull distributions~\cite{KLWeibull-2013} is
\begin{equation}
D_\KL(p_{k_1,s_1}:p_{k_2,s_2})=
\log \frac{k_{1}}{s_{1}^{k_{1}}}
-\log \frac{k_{2}}{s_{2}^{k_{2}}}
+\left(k_{1}-k_{2}\right)\left[\log s_{1}-\frac{\gamma}{k_{1}}\right]+\left(\frac{s_{1}}{s_{2}}\right)^{k_{2}} \Gamma\left(\frac{k_{2}}{k_{1}}+1\right)-1. \label{eq:KLWeibull}
\end{equation}

We check that the KLD between two scale Weibull families is scale invariant:
\begin{equation}
\forall \lambda>0,\quad D_\KL(p_{k_1,\lambda s_1}:p_{k_2,\lambda s_2})=D_\KL(p_{k_1,s_1}:p_{k_2,s_2}),
\end{equation}
and that
\begin{equation}
 D_\KL(p_{k_1,s_1}:p_{k_2,s_2})= 
D_\KL(p_{k_1}:p_{k_2,\frac{s_2}{s_1}})= D_\KL(p_{k_1,\frac{s_1}{s_2}}:p_{k_2}).
\end{equation}

Indeed, we can rewrite equivalently Eq.~\ref{eq:KLWeibull} as:
\begin{equation}
D_\KL(p_{k_1,s_1}:p_{k_2,s_2})=  
\left(\frac{s_{1}}{s_{2}}\right)^{k_{2}} \Gamma\left(\frac{k_{2}}{k_{1}}+1\right) 
-k_2\log\frac{s_1}{s_2}
+ \log\frac{k_1}{k_2}
-\left(1-\frac{k_2}{k_1}\right)\gamma -1.
\end{equation}
This last expression highlights the use of the scale invariant ratio $\lambda=\frac{s_1}{s_2}$.

When $k_1=k_2=k$, the KLD between two Weibull densities of $\calF_{p_k}$ is:
\begin{equation}
D_\KL(p_{k,s_1}:p_{k,s_2})= \left(\frac{s_1}{s_2}\right)^k-k\log\frac{s_1}{s_2}-1,
\end{equation}
since $\Gamma(2)=1$. 
In that case, since $\calF_{p_k}$ is an exponential family, we check 
that in the case the KLD amounts to the Itakura-Saito divergence (a Bregman divergence) on 
the swapped natural parameter $\theta_i=\frac{1}{s^k_i}$.

The KLD between an exponential distribution ($k_1=1$) and a Rayleigh distribution ($k_2=2$) is
\begin{eqnarray}
D_\KL(p_{s_1}^\calE:p_{s_2}^\calR) &=& 2\left(\frac{s_1}{s_2}\right)^2-\log \left(\frac{s_1}{s_2}\right)^2 +c,\\
&=& 2\lambda^2-2\log\lambda+c
\end{eqnarray}
since $\Gamma(2+1)=2$, and where $c$ denotes a constant.
It follows that $D_\KL(p^\calE:p_{s}^\calR)=\frac{2}{s^2}-\log\frac{1}{s^2}+c$ is minimized for $s=\sqrt{2}$.
Conversely, $D_\KL(p_s^\calE:p^\calR)=2s^2-\log s^2+c$ is minimized for $s=\frac{1}{\sqrt{2}}$.


The exponential and Rayleigh scale families are 1D submanifolds of the Weibull manifold whose information-geometrc structure has been studied in~\cite{WeibullIG-2008}.
\end{example}

Last but not least, these results apply for families of distributions $p_X$ that can be transformed into a location-scale family via an invertible and differentiable transformation (e.g., example~\ref{ex:lognormal}).

\appendix

\section{Fisher-Rao distance between two densities of a location-scale family}\label{sec:frls}

Let $\calF_p=\left\{ p_{l,s}(x):=\frac{1}{s}p\left(\frac{x-l}{s}\right) \st (l,s)\in \bbR\times\bbR_{++}\right\}$ 
be a location-scale family induced by the standard density $p(x)$ with support $\calX=\bbR$. 
Location-scale families include the family of normal distributions, the family of Laplace distributions, the family of Student $t$-distributions (including the family of Cauchy distributions), the family of logistic distributions, the families of univariate elliptical distributions~\cite{Mitchell-1988}, etc.

Under mild regularity conditions (i.e., interchanging derivation and integration operation order),
 the Fisher information matrix (FIM) $I_\lambda(\lambda)$ with respect to parameter $\lambda=(l,s)\in\bbR\times\bbR_{++}$ is given by:
\begin{eqnarray}
I_\lambda(\lambda) &=& E_{p_\lambda}\left[\nabla_\lambda \log p_\lambda(x) (\nabla_\lambda \log p_\lambda(x))^\top\right],\\
&=& - E_{p_\lambda}\left[\nabla_\lambda^2 \log p_\lambda(x)\right].
\end{eqnarray}

Let $g_{ij}(\lambda$ denote the $(i,j)$-th coefficient of the FIM so that we have $I_\lambda(\lambda)=[g_{ij}(\lambda)]_{ij}$ with
\begin{eqnarray}
g_{ij}(\lambda) &=& E_{p_\lambda}[\partial_i \log p_\lambda(x) \partial_j \log p_\lambda(x)],\\ 
&=& -E_{p_\lambda} \left[\partial_i\partial_j \log p_\lambda(x)\right],
\end{eqnarray}
where $\partial_i :=\frac{\partial}{\partial \lambda_i}$.

When handling location-scale densities $p_{l,s}(x):=\frac{1}{s}p\left(\frac{x-l}{s}\right)$, we shall observe that using a change of variable $y=\frac{x-l}{s}$ (with $\dy=\frac{\dx}{s}$), we have for any function $f$ the following identity: 
\begin{eqnarray}
E_{p_\lambda}\left[f\left(\frac{x-l}{s}\right)\right] &=&\int \frac{1}{s} p\left(\frac{x-l}{s}\right)f\left(\frac{x-l}{s}\right) \dx ,\\
&=& \int p(y)f(y)\dy = E_p[f(x)].
\end{eqnarray}

The log-likelihood of a location-scale density is $\log p_{l,s}(x)=\log p\left(\frac{x-l}{s}\right)-\log s$.
Let us compute the coefficients of the FIM using the notations $\partial_l =\frac{\partial}{\partial l}$ and $\partial_s =\frac{\partial}{\partial s}$ as follows:

\begin{itemize}
\item Let us compute the first diagonal coefficient of the FIM using
\begin{equation}
\partial_l \log p_{l,s}(x) = -\frac{1}{s} \frac{p'\left(\frac{x-l}{s}\right)}{p\left(\frac{x-l}{s}\right)},
\end{equation}
so that it comes that:
\begin{eqnarray}
g_{11}(\lambda) &=& E_{p_\lambda}\left[(\partial_l \log p_{l,s}(x) )^2\right],\\
&=& \frac{1}{s^2} E_{p_\lambda}\left[ \left(\frac{p'\left(\frac{x-l}{s}\right)}{p\left(\frac{x-l}{s}\right)}\right)^2\right],\\
&=& \frac{1}{s^2} E_{p}\left[\left(\frac{p'\left(x\right)}{p\left(x\right)}\right)^2\right].
\end{eqnarray}

\item We proceed and compute the second diagonal coefficient of the FIM using
\begin{eqnarray}
\partial_s \log p_{l,s}(x) &=& -\frac{1}{s^2} (x-l) \frac{p'\left(\frac{x-l}{s}\right)}{p\left(\frac{x-l}{s}\right)}  -\frac{1}{s},\\
&=& -\frac{1}{s} \left(1+ \frac{x-l}{s}\frac{p'\left(\frac{x-l}{s}\right)}{p\left(\frac{x-l}{s}\right)}  \right),
\end{eqnarray}
so that it comes that
\begin{eqnarray}
g_{22}(\lambda) &=& E_{p_\lambda}\left[(\partial_s \log p_{l,s}(x) )^2\right],\\
&=& \frac{1}{s^2} E_{p_\lambda}\left[\left(1+ \frac{x-l}{s}\frac{p'\left(\frac{x-l}{s}\right)}{p\left(\frac{x-l}{s}\right)}  \right)^2\right],\\
&=& \frac{1}{s^2} E_{p}\left[\left(1+ x \frac{p'(x)}{p(x)}\right)^2\right].
\end{eqnarray}

\item Finally, we compute the off-diagonal coefficients of FIM as follows:
\begin{eqnarray}
g_{12}(\lambda) &=&  g_{21} = E_{p_\lambda}\left[(\partial_l \log p_{l,s}(x))(\partial_s \log p_{l,s}(x))  \right],\\
&=& E_{p_\lambda}\left[(\partial_l \log p_{l,s}(x))(\partial_s \log p_{l,s}(x))  \right],\\
&=& \frac{1}{s^2} E_{p_\lambda}\left[  \frac{p'\left(\frac{x-l}{s}\right)}{p\left(\frac{x-l}{s}\right)} 
 \left(1+ \frac{x-l}{s}\frac{p'\left(\frac{x-l}{s}\right)}{p\left(\frac{x-l}{s}\right)}  \right) \right],\\
&=& \frac{1}{s^2} E_{p}\left[ \frac{p'(x)}{p(x)} \left(1+ x \frac{p'(x)}{p(x)} \right) \right].
\end{eqnarray}

\end{itemize}

Thus the FIM of a location-scale family with respect to parameter $\lambda=(l,s)$ writes as follows
\begin{equation}
I_\lambda(\lambda)=\frac{1}{s^2}\mattwotwo{a^2}{c}{c}{b^2},
\end{equation}
with the following constants depending on the standard density $p$:
\begin{eqnarray}
a^2 &=& E_p\left[\left(\frac{p'(x)}{p(x)}\right)^2\right]\geq 0,\\
b^2 &=& E_p\left[\left(1+x\frac{p'(x)}{p(x)}\right)^2\right]\geq 0,\\
c &=& E_p\left[\frac{p'(x)}{p(x)} \left(1+ x \frac{p'(x)}{p(x)} \right) \right].
\end{eqnarray}

\begin{BoxedProposition}[Fisher information of a location-scale family]
The Fisher information matrix $I(\lambda)$ of a location-scale family with continuously differentiable standard density $p(x)$ with full support $\bbR$ is
 $I(\lambda)=\frac{1}{s^2}\mattwotwo{a^2}{c}{c}{b^2}$, where 
$a^2= E_p\left[\left(\frac{p'(x)}{p(x)}\right)^2\right]$, $b^2= E_p\left[\left(1+x\frac{p'(x)}{p(x)}\right)^2\right]$
 and $c = E_p\left[\frac{p'(x)}{p(x)} \left(1+ x \frac{p'(x)}{p(x)} \right) \right]$.
\end{BoxedProposition}

Note that when $c\not=0$, the parameters $l$ and $s$ are correlated (i.e., non-orthogonal).
Assume the standard density is an even function (e.g., the normal, Cauchy, and Laplace standard densities):
We have $p(-x)=p(x)$ and its derivative $p'(x)$ is odd: $p'(-x)=-p'(x)$.  
Then the function $h(x)=\frac{p'(x)}{p(x)} \left(1+ x \frac{p'(x)}{p(x)}\right)$ is odd since $\frac{p'(x)}{p(x)}$ is odd and 
$\left(1+ x \frac{p'(x)}{p(x)}\right)$ is even.
We have $E_p[h(x)]=0$ for any odd function $h(x)$ and even density $p(x)$:
Indeed, by a change of variable $y=-x$ in the integral $\int_{-\infty}^0 \ldots\dx$, we find that
\begin{eqnarray}
E_p[h(x)] &=& \int_{-\infty}^\infty p(x)h(x)\dx,\\
&=& \int_{-\infty}^0 p(x)h(x)\dx +\int_{0}^\infty p(x)h(x)\dx,\\
&=&  \int_{+\infty}^0 p(y)h(y)\dy +\int_{0}^\infty p(x)h(x)\dx,\\
&=& -\int_{0}^\infty p(x)h(x)\dx+\int_{0}^\infty p(x)h(x)\dx,\\
&=& 0.
\end{eqnarray}

Notice that even standard density $p(x)$ are symmetric and have zero skewness $E_p\left[x^3\right]$ since $x^3$ is an odd function.

Thus let us consider that the standard density is an even function so that the FIM 
with respect to parameter $\lambda=(l,s)$ is the following diagonal matrix:
\begin{equation}
I_\lambda(\lambda)=\frac{1}{s^2}\mattwotwo{a^2}{0}{0}{b^2},
\end{equation}
with
\begin{eqnarray}
a^2 &=& E_p\left[\left(\frac{p'(x)}{p(x)}\right)^2\right]\geq 0,\\
b^2 &=& E_p\left[\left(1+x\frac{p'(x)}{p(x)}\right)^2\right]>0.
\end{eqnarray}

Furthermore, let us reparameterize the location-scale density by $\theta(\lambda)=\left(\frac{a}{b}\lambda_1,\lambda_2\right)$ where $a=\sqrt{a^2}$ and $b=\sqrt{b^2}$ are the positive square roots of $a^2$ and $b^2$, respectively.
We have $\lambda(\theta)=\left(\frac{b}{a}\theta_1,\theta_2\right)$.

Using the covariance transformation of the FIM~\cite{EIG-2020}, we get
\begin{eqnarray}
I_\theta(\theta) &=& \left[\frac{\partial\lambda_i}{\partial\theta_j} \right]^\top_{ij} \times  I_\lambda(\lambda(\theta)) \times \left[\frac{\partial\lambda_i}{\partial\theta_j} \right]_{ij},\\
&=& \mattwotwo{\frac{b}{a}}{0}{0}{1} \times  \mattwotwo{\frac{a^2}{\theta_2^2}}{0}{0}{\frac{b^2}{\theta_2^2}} \times \mattwotwo{\frac{b}{a}}{0}{0}{1},\\
&=& \frac{b^2}{\theta_2^2}\, \mattwotwo{1}{0}{0}{1}.
\end{eqnarray}

This metric corresponds to a scaled metric of the Poincar\'e upper plane (conformal metric).
It follows that the Gaussian curvature $\kappa$ is constant and negative: 
\begin{equation}
\kappa=-\frac{1}{b^2}<0.
\end{equation}

Thus the Fisher-Rao distance between two densities of a location-scale family is hyperbolic. 
Let $\rho_{U,\kappa}$ denote the hyperbolic distance in the hyperbolic geometry of curvature $\kappa$~\cite{HVD-2010,HVD-2012}:
\begin{equation}
\rho_{U,\kappa}(\theta_1,\theta_2)=\sqrt{-\frac{1}{\kappa}}  \arccosh\left(  \frac{1-\theta_1\cdot\theta_2}{\sqrt{(1-\theta_1\cdot\theta_1)(1-\theta_2\cdot\theta_2)}}  \right)
,
\end{equation}
where $\arccosh(u)=\log(u+\sqrt{u^2-1})$ for $u>1$ and $\cdot$ denotes the scalar product: $\theta\cdot\theta'=\theta^\top\theta'=\theta_1\theta_1'+\theta_2\theta_2'$.

Thus we get the following proposition:

\begin{BoxedProposition}[Fisher-Rao distance on a location-scale manifold]
The Fisher-Rao distance between two densities $p_{l_1,s_1}$ and $p_{l_2,s_2}$ of a location-scale family $\calF_p$ with even standard density $p(x)=p(-x)$ on the support $\bbR$ is
$$ 
\rho_{p}((l_1,s_1),(l_2,s_2))= b\ \rho_U\left(\left(\frac{a}{b}l_1,s_1\right),\left(\frac{a}{b}l_2,s_2\right)\right),
$$ 
where $a=\sqrt{E_p\left[\left(\frac{p'(x)}{p(x)}\right)^2\right]}$ and $b=\sqrt{E_p\left[x\left(\frac{p'(x)}{p(x)}+1\right)\right]}>0$, and
$$ 
\rho_U((l_1,s_1),(l_2,s_2))=\arccosh\left(  \frac{1-(l_1l_2+s_1s_2)}{\sqrt{\left(1-\left(l_1^2+s_1^2\right)\right)\left(1-\left(l_2^2+s_2^2\right)\right)}}  \right).
$$ 
\end{BoxedProposition}

\begin{example}
The Fisher-Rao distance between two normal densities $p_{\mu_1,\sigma_1}^\calN$ and $p_{\mu_12,\sigma_2}^\calN$ is
\begin{equation}
\rho_{p^\calN}((\mu_1,\sigma_1),(\mu_2,\sigma_2))=
\sqrt{2}\ 
\rho_U\left(\left(\frac{1}{\sqrt{2}}\mu_1,\sigma_1\right),\left(\frac{1}{\sqrt{2}}\mu_2,\sigma_2\right)\right)
\end{equation}
since $a^2=1$, $b^2=2$, $\kappa=-\frac{1}{2}$.
\end{example}

\begin{example}
The Fisher-Rao distance between two Cauchy densities is a scaled hyperbolic distance
\begin{equation}
\rho_{p^\calC}((l_1,s_1),(l_2,s_2))= \frac{1}{\sqrt{2}}\ \rho_U\left((l_1,s_1),(l_2,s_2)\right),
\end{equation}
since $a^2=b^2=\frac{1}{2}$ and $\kappa=-\frac{1}{b^2}=-2$.
\end{example}


Consider the mapping $(l,s)\mapsto \frac{a}{b}l+is\in\bbC$ where $i^2=-1$.
The Poincar\'e complex upper plane $\bbU$ can be transformed into the Poincar\'e complex disk $\bbD$ using a Cayley transform~\cite{HVD-2010,HilbertSiegel-2020}.
Let $\SL_\bbR(2)$  be the group represented by the matrices:
\begin{equation}
\SL_\bbR(2) := \left\{ \mattwotwo{a}{b}{c}{d} \st  a,b,c,d\in\bbR,\quad ad-bc=1\right\}.
\end{equation}
The action of the group $\SL_\bbR(2)$ on $\bbU$ is defined by real linear fractional transforms (M\"obius transformations) $z\mapsto \frac{az+b}{cz+d}$ for $\mattwotwo{a}{b}{c}{d} \in\SL_\bbR(2)$ defined on the extended complex plane $\bbC\cup\{\infty\}$.
Let $\SU_\bbC(1,1)$  denote the special unitary group:
\begin{equation}
\SU_\bbC(1,1) :=\left\{ \mattwotwo{a}{b}{\bar{b}}{\bar{a}} \st  a,b\in\bbC,\quad a\bar{a}-b\bar{b}=1  \right\}.
\end{equation}
The group $\SU_\bbC(1,1)$ acts on $\bbD$ via complex linear fractional transforms: $z\mapsto \frac{az+b}{\bar{b}z+\bar{a}}$.
Notice that the group $\SL_\bbR(2)$ is isomorphic to group $\SU_\bbC(1,1)$:
Using the matrix representations, we have $A\in \SL_\bbR(2)\mapsto CAC^{-1}\in \SU_\bbC(1,1)$ where $C=\mattwotwo{1}{-i}{1}{i}$.
Thus we can convert $\bbU$ to $\bbD$ using the transformation $\frac{z-i}{z+i}$, and reciprocally we convert $\bbD$ to $\bbU$ using the inverse transformation $\frac{i(z+1)}{1-z}$.
When performing geometric computing, it is preferable to use the Klein model of hyperbolic geometry since geodesics are straight lines restricted to the open unit disk.


%
%
%
\small
\bibliographystyle{plain}
\bibliography{KLCauchyBIBV3}

\begin{thebibliography}{10}

\bibitem{IG-2016}
Shun-ichi Amari.
\newblock {\em Information Geometry and Its Applications}.
\newblock Applied Mathematical Sciences. Springer Japan, 2016.

\bibitem{EF-2014}
Ole Barndorff-Nielsen.
\newblock {\em Information and exponential families}.
\newblock John Wiley \& Sons, 2014.

\bibitem{StatTransfModel-2012}
Ole~E Barndorff-Nielsen, Preben Bl{\ae}sild, and Poul~S Eriksen.
\newblock {\em Decomposition and invariance of measures, and statistical
  transformation models}, volume~58.
\newblock Springer Science \& Business Media, 2012.

\bibitem{KLWeibull-2013}
Christian Bauckhage.
\newblock {Computing the Kullback-Leibler Divergence between two Weibull
  Distributions}.
\newblock {\em arXiv preprint arXiv:1310.3713}, 2013.

\bibitem{BD-1967}
Lev~M Bregman.
\newblock The relaxation method of finding the common point of convex sets and
  its application to the solution of problems in convex programming.
\newblock {\em USSR computational mathematics and mathematical physics},
  7(3):200--217, 1967.

\bibitem{WeibullIG-2008}
Limei Cao, Huafei Sun, and Xiaojie Wang.
\newblock The geometric structures of the {W}eibull distribution manifold and
  the generalized exponential distribution manifold.
\newblock {\em Tamkang Journal of Mathematics}, 39(1):45--51, 2008.

\bibitem{LogNormal-1987}
Edwin~L Crow and Kunio Shimizu.
\newblock {\em Lognormal distributions}.
\newblock Marcel Dekker New York, 1987.

\bibitem{Csiszar-1967}
Imre Csisz{\'a}r.
\newblock Information-type measures of difference of probability distributions
  and indirect observation.
\newblock {\em studia scientiarum Mathematicarum Hungarica}, 2:229--318, 1967.

\bibitem{InfoProj-1975}
Imre Csisz{\'a}r.
\newblock {$I$-divergence geometry of probability distributions and
  minimization problems}.
\newblock {\em The annals of probability}, pages 146--158, 1975.

\bibitem{Iprojection-1984}
Imre Csisz{\'a}r.
\newblock {Sanov property, generalized $I$-projection and a conditional limit
  theorem}.
\newblock {\em The Annals of Probability}, pages 768--793, 1984.

\bibitem{Csiszar-2003}
Imre Csisz{\'a}r and Frantisek Matus.
\newblock Information projections revisited.
\newblock {\em IEEE Transactions on Information Theory}, 49(6):1474--1490,
  2003.

\bibitem{MVNClustering-2007}
Jason~V Davis and Inderjit~S Dhillon.
\newblock Differential entropic clustering of multivariate {G}aussians.
\newblock In {\em Advances in Neural Information Processing Systems}, pages
  337--344, 2007.

\bibitem{Renyi-2013}
Manuel Gil, Fady Alajaji, and Tamas Linder.
\newblock R{\'e}nyi divergence measures for commonly used univariate continuous
  distributions.
\newblock {\em Information Sciences}, 249:124--131, 2013.

\bibitem{Hayashi-2011}
Masahito Hayashi.
\newblock Large deviation theory for non-regular location shift family.
\newblock {\em Annals of the Institute of Statistical Mathematics},
  63(4):689--716, 2011.

\bibitem{Hotelling-1930}
Harold Hotelling.
\newblock Spaces of statistical parameters.
\newblock {\em Bull. Amer. Math. Soc}, 36:191, 1930.

\bibitem{KassVos-2011}
Robert~E Kass and Paul~W Vos.
\newblock {\em Geometrical foundations of asymptotic inference}, volume 908.
\newblock John Wiley \& Sons, 2011.

\bibitem{StatMatrices-2006}
Tonu Kollo and Dietrich von Rosen.
\newblock {\em Advanced multivariate statistics with matrices}, volume 579.
\newblock Springer Science \& Business Media, 2006.

\bibitem{komaki-2007}
Fumiyasu Komaki.
\newblock Bayesian prediction based on a class of shrinkage priors for
  location-scale models.
\newblock {\em Annals of the Institute of Statistical Mathematics},
  59(1):135--146, 2007.

\bibitem{DensityMfd-1988}
John~D Lafferty.
\newblock The density manifold and configuration space quantization.
\newblock {\em Transactions of the American Mathematical Society},
  305(2):699--741, 1988.

\bibitem{Lax-2001}
Peter~D Lax.
\newblock {Change of variables in multiple integrals II}.
\newblock {\em The American Mathematical Monthly}, 108(2):115--119, 2001.

\bibitem{Mitchell-1989}
Ann~ES Mitchell.
\newblock The information matrix, skewness tensor and $\alpha$-connections for
  the general multivariate elliptic distribution.
\newblock {\em Annals of the Institute of Statistical Mathematics},
  41(2):289--304, 1989.

\bibitem{Mitchell-1988}
Ann~FS Mitchell.
\newblock Statistical manifolds of univariate elliptic distributions.
\newblock {\em International Statistical Review/Revue Internationale de
  Statistique}, pages 1--16, 1988.

\bibitem{SIMBAD-2013}
Frank Nielsen.
\newblock Pattern learning and recognition on statistical manifolds: an
  information-geometric review.
\newblock In {\em International Workshop on Similarity-Based Pattern
  Recognition}, pages 1--25. Springer, 2013.

\bibitem{InformationProjection-2018}
Frank Nielsen.
\newblock What is... an information projection?
\newblock {\em Notices of the AMS}, 65(3):321--324, 2018.

\bibitem{EIG-2020}
Frank Nielsen.
\newblock An elementary introduction to information geometry.
\newblock {\em Entropy}, 22(10):1100, 2020.

\bibitem{CauchyVoronoi-2020}
Frank Nielsen.
\newblock On {V}oronoi diagrams on the information-geometric {C}auchy
  manifolds.
\newblock {\em Entropy}, 22(7):713, 2020.

\bibitem{HilbertSiegel-2020}
Frank Nielsen.
\newblock {The Siegel--Klein Disk: Hilbert Geometry of the Siegel Disk Domain}.
\newblock {\em Entropy}, 22(9):1019, 2020.

\bibitem{EF-2009}
Frank Nielsen and Vincent Garcia.
\newblock {Statistical exponential families: A digest with flash cards}.
\newblock {\em arXiv preprint arXiv:0911.4863}, 2009.

\bibitem{HVD-2010}
Frank Nielsen and Richard Nock.
\newblock {Hyperbolic Voronoi diagrams made easy}.
\newblock In {\em 2010 International Conference on Computational Science and
  Its Applications}, pages 74--80. IEEE, 2010.

\bibitem{HVD-2012}
Frank Nielsen and Richard Nock.
\newblock {The hyperbolic Voronoi diagram in arbitrary dimension}.
\newblock {\em arXiv preprint arXiv:1210.8234}, 2012.

\bibitem{NN-2013}
Richard Nock, Brice Magdalou, Eric Briys, and Frank Nielsen.
\newblock Mining matrix data with {B}regman matrix divergences for portfolio
  selection.
\newblock In {\em Matrix Information Geometry}, pages 373--402. Springer, 2013.

\bibitem{Ohara-2014}
Atsumi Ohara and Shinto Eguchi.
\newblock {Geometry on positive definite matrices deformed by $V$-potentials
  and its submanifold structure}.
\newblock In {\em Geometric Theory of Information}, pages 31--55. Springer,
  2014.

\bibitem{NIG-1995}
Giovanni Pistone, Carlo Sempi, et~al.
\newblock An infinite-dimensional geometric structure on the space of all the
  probability measures equivalent to a given one.
\newblock {\em The annals of statistics}, 23(5):1543--1561, 1995.

\bibitem{fdivdiffeo-2010}
Yu~Qiao and Nobuaki Minematsu.
\newblock A study on invariance of $f$-divergence and its application to speech
  recognition.
\newblock {\em IEEE Transactions on Signal Processing}, 58(7):3884--3890, 2010.

\bibitem{MC-2013}
Christian Robert and George Casella.
\newblock {\em {Monte Carlo statistical methods}}.
\newblock Springer Science \& Business Media, 2013.

\bibitem{WarpedRieMetricsLS-2019}
Salem Said, Lionel Bombrun, and Yannick Berthoumieu.
\newblock {Warped Riemannian metrics for location-scale models}.
\newblock In {\em Geometric Structures of Information}, pages 251--296.
  Springer, 2019.

\bibitem{Skovgaard-1984}
Lene~Theil Skovgaard.
\newblock A {R}iemannian geometry of the multivariate normal model.
\newblock {\em Scandinavian journal of statistics}, pages 211--223, 1984.

\bibitem{Villa-2016}
Cristiano Villa.
\newblock {A property of the Kullback--Leibler divergence for location-scale
  models}.
\newblock {\em arXiv preprint arXiv:1604.01983}, 2016.

\bibitem{NIG-2013}
Jun Zhang.
\newblock Nonparametric information geometry: From divergence function to
  referential-representational biduality on statistical manifolds.
\newblock {\em Entropy}, 15(12):5384--5418, 2013.

\end{thebibliography}

\end{document}